\documentclass[12pt]{article} 
\usepackage{setspace} 
\usepackage{amsmath}
\usepackage{amsfonts}
\usepackage{amssymb}
\usepackage{graphicx}

\title{Estimating the theoretical error rate for prediction}
\author{Herman Chernoff$^1$, Shaw-Hwa Lo$^2$, Tian Zheng$^2$, Adeline Lo$^3$}
\date{$^1$Department of Statistics, Harvard University, Cambridge, MA. \\$^2$Department of Statistics, Columbia University, New York, NY\\ $^3$Department of Politics, Princeton University, Princeton, NJ}

\begin{document}

\maketitle

\begin{abstract}
Prediction for very large data sets is typically carried out in two stages, 
variable selection and pattern recognition. Ordinarily
variable selection involves seeing how well individual explanatory  
variables are correlated with the dependent variable. This practice  
neglects the possible interactions among
the variables. Simulations have shown that a statistic $I$, 
that we used for variable selection is much better correlated 
with predictivity than significance
levels. We explain this by defining  theoretical predictivity and show
how $I$ is related to predictivity. We calculate the biases
of the overoptimistic {\it training estimate} of predictivity and of the
pessimistic {\it out of sample} estimate. Corrections for the bias
lead to improved estimates of the potential predictivity  using small
groups of possibly interacting variables. These results support the use
of $I$ in the variable selection phase  of prediction for
data sets such as in GWAS (Genome wide association studies) where there
are very many explanatory variables and modest sample sizes.   
Reference is made to another publication using $I$, which led 
to a reduction in the error rate of prediction from
30\% to 8\%, for a data set with, 4,918 variables
and 97 subjects. This data set had been previously studied 
by scientists for over 10 years.

\end{abstract}

\section{INTRODUCTION}
The problems of prediction and classification $(C)$ have a long history in 
the statistics literature. Recent advances in technology confront us with 
problems with which the classical literature did not deal, and new
approaches are necessary and are being tried. An article in Nature
Genetics [1], ``Predicting the influence of common variants", identified
prediction as an important additional goal  for current genome-wide 
association studies $(GWAS)$. A common approach consists of the two
parts, Variable Selection $(VS)$, in which a few highly relevant explanatory
variables are selected from the many that are available, 
and $C$,  applying pattern recognition 
techniques on these variables to make predictions
for new subjects. Ordinarily $VS$ selects variables by how well they are
correlated with the outcome. Recently an article [2] pointed out that adding
highly significant variables to a group does not necessarily increase
predictivity. A recent report in PNAS [3] pointed out that significance is
not necessarily well related to predictivity, suggesting that  predictivity
requires a new framework, and that a statistic $I$ that was used in a method
of $VS$ called {\it Partition Retention} $(PR)$ [4] might be a useful tool in
prediction since, in simulations, it was much better correlated with
predictivity than was significance.

This paper confronts some of the issues in variable selection for prediction
with large data sets. In the interest of simplicity, we shall confine 
ourselves to a special problem, but the concepts  are much more
generally applicable, and may serve as a framework
for a theory. This problem is a case-control study  that
involves two states, $h$ for Healthy and $d$
for Disease. and a sample of $n$ subjects for each.  There are
a large number, $m$, of explanatory variables {\it markers} called SNPs
for each subject. In some current studies $n$ may
vary from a hundred to a few thousand while $m$ may vary from several
hundred to a million. The object of these
studies may be to find out
which of the variables influence the disease or, to use knowledge of these
variables for another individual, to predict whether he has the disease.

 Prediction ordinarily requires some variable selection, but
the scientist who wants to understand how the disease works may not be
primarily interested in prediction and may have somewhat different criteria 
for variable selection than the predictor.  Once the predictor has selected
a subset of the variables, he still faces a substantial
pattern recognition problem of deciding
on a strategy for using them to make his predictions. It would be helpful
to have an estimate of how well he can hope to do with this subset.
If he falls far short of this estimate, there may be room for improvement.

A standard approach to the variable selection problem is to see how well
each variable is correlated with the outcome, and to select a few that
are most correlated.  One need not depend on linear
correlation. One can simply measure the significance level for the test
that the two distributions of the variable under $h$ and $d$ are the same.
A $t$ test will provide a p-value which can be used to compare the various
one dimensional candidate variables.

This approach will work well for a simple disease where certain variables have a great influence on the disease by themselves. But if the disease is 
influenced by the interaction among several variables, 
none of the influential
variables may show up as significant, whereas if we have 
a million candidate variables,
some noninfluential ones may show up as very significant by chance alone.

There is an implicit feeling that the variables that will be useful
for $VS$ and $C$ will show up as highly significant in testing the null
hypothesis that the two  distributions under $h$ and $d$  are the same.
When we consider a group of two or more variables, a typical test of the
null hypothesis may be something like a chi-square test, which is
not as focused as the $t$ test in the one dimensional problem, and may lack
power. The significance level obtained by this test may be a poor 
indicator of the classification power of the group of variables.
Moreover, 
the computational burden, of dealing with the interaction of only two
of the variables at a time, can be overwhelming when $m \approx 1,000,000$.

The technique for $VS$, {\it Partition Retention} $(PR)$,
copes with the computational problem involved in large data sets
where there may be interaction among the influential explanatory variables.
Wang et al [5] attacked a data set with 4,918 gene expression variables and 97 subjects using $PR$ to reduce the set of variables to
a few overlapping small groups of
interacting variables, and then worked on the pattern
recognition problem. They were successful in reducing a standard error rate
of about 30\% to 8\% on this data set that had been studied by many
scientists for over 10 years. 

From the prediction point of view, there is even the possibility that two
different variables are influential  and may show up to be significant,
but each only influences a small percentage of subjects. Then prediction
based on these variables will work only for a small fraction of cases. 
In short, the p-value for testing the null hypothesis, using a group of       
variables,  is an unreliable sign of how useful the group will be for
$C$ and we have conjectured that $I$ may be a useful measure in selecting
variables for $C$.

When the number of explanatory variables is very large, 
the possibility of a satisfactory
resolution of the $VS$ and $C$ problems depends on the underlying simplicity
of the situation. If the real relationships are very complicated, it may 
not be possible to find an adequate solution.

The object of this paper is to show how the {\it statistic} $I$, that plays
a fundamental role in the $PR$ method, is related to a 
{\it parameter} $\theta_e$ that
measures the potential ability to use a small group of explanatory 
variables for classification.
Since $\theta_e$ tells us how well we can hope to do, we discuss several
estimates of $\theta_e$, their biases and corrections for those biases.
We show that Partition Retention for VS can be a useful tool in developing
good prediction procedures.

\section{THE PARAMETER FOR $C$}
In a case control study the statistician who is given a small 
group $X$ of discrete valued explanatory variables 
obtains data with which to estimate the underlying distributions 
of $X$ under $d$ for Disease and $h$ for Health. The 
ability, to use the data of this group for classification, depends
on these probability distributions, $f_d$ and $f_h$,
which the statistician can only estimate. 

Whereas calculating significance involves using the case control data
to test  whether $f_d=f_h$,  prediction involves the use of  
the observation $X$ on an individual for 
testing whether it comes from $f_d$ or from $f_h$.
Suppose that we know
these probability distributions and we consider $d$ and $h$ as equally
likely {\it a priori}, and regard as equally bad,
the two possible errors of a false positive and a false negative.
Then the best decision rule (prediction) for a possible
observation $x$ on $X$ is to select according to the  greater of $f_d(x)$ and
$f_h(x)$. The probability of a false positive is
$Pr(f_d(X)>f_h(X)|f_h)=\sum_{x:f_d(x)>f_h(x)} f_h(x)$ and  
the probability of a 
false negative is
$Pr(f_d(X)\leq f_h(X)|f_d)=\sum_{x:f_d(x)\leq f_h(x)} f_d(x).$

Then the ideal average error rate and 
correct classification rate for this procedure based on information that
the statistician lacks, but can estimate,
are 
\begin{equation}
\theta_e=0.5\sum_x \min(f_d(x),f_h(x))
\end{equation}
and $\theta_c=1-\theta_e=0.5\sum_x \max(f_d(x),f_h(x))$. Taking 
the difference of the two sums, we can write 
\begin{equation}
\sum_x|f_d(x)-f_h(x)|=2\theta_c-2\theta_e=2-4\theta_e
\end{equation}
which represents $\theta_c$ and $\theta_e$ in terms of the sum of 
the absolute differences of the probability densities. A similar
framework with variable selection can be found in [6].

In  prediction we ordinarily have different prior probabilities and different
costs of error. These suggest that we later consider a 
natural modification of this definition of
the underlying parameter representing the ability to predict.

One may use the data, {\it i.e.} the training set, to estimate $\theta_e$.
The naive {\it training estimate} of $\theta_e$ is
\begin{equation}
 {\hat \theta}_e = 0.5\sum_x\min(n_{dx},n_{hx})/n,
\end{equation} 
where $n_{dx}$ and $n_{hx}$ are the number of diseased and healthy
observations for a  given value $x$. This {\it training estimate} 
of $\theta_e$ tends 
to be overoptimistic.  Note that it can
be expressed in terms of $\sum_x|n_{dx}-n_{hx}|$.

Note also that $\theta_e$ is an ideal and the statistician, given the
data and not $f_d$ and $f_h$, may lack a good strategy to attain 
something  close  to that
error rate. Bounds on $\theta_e$ would be useful in deciding whether a
method of prediction is adequate or requires enhancement.

\section{$I$ SCORE, BOUNDS, ESTIMATORS AND BIASES}

In an intensive calculation from [3]  involving 6 snps with specified minor
allele frequencies (MAF) and specified probabilities of response, it was
indicated that when $\theta_c$ was known, neither significance levels nor
training prediction rates were
well correlated with this parameter unless the sample size was
very large, while $I$ seemed well correlated with it for quite
moderate sample sizes. This paper explains that phenomenon,

\subsection{The Score $I$}
We present a
definition of $I$ when we have a sample of $n_0$ observations on
$(Y,X)$ where $Y$ is a dependent variable which has been normalized to have
sample mean 0 and sample variance 1,
and $X$ represents a discrete valued variable  
or a group of discrete valued explanatory variables. 
\begin{equation}
I = n_0^{-1}\sum_x(n_x{\bar Y}_x)^2 .
\end{equation}
Here $n_x$ is the number of observations where $X=x$, ${\bar Y}_x$ is
the sample mean of the values of $Y$ for which $X=x$. 

In our application, we 
identify $d$ and $h$ with $Y=1$ and $Y=-1$. Because we have equal sample
sizes the sample mean ${\bar Y}=0$ and the 
sample variance, $n_0^{-1}\sum (Y-{\bar Y})^2=1$, and
 $Y$ is automatically normalized. Also 
$n_x=n_{dx}+n_{hx},\,n_0=2n$ and $n_x{\bar Y}_x = n_{dx}-n_{hx}$. Thus
\begin{equation}
 I = 0.5\,n^{-1}\sum_x(n_{dx}-n_{hx})^2,
\end{equation} 
which seems related to the training prediction rate. For large
samples, we may think of $I/n$ as an estimate of the parameter
\begin{equation}
\theta_I= 0.5\sum_x(f_d(x)-f_h(x))^2 .
\end{equation}
To study the expectations and variances of $I$ and the 
training prediction rate, we should be  interested in the 
moments of $|n_{dx}-n_{hx}|$.

It should be noted that $I$ is a modified version of 
$$J=n^{-1}\sum_x n_x{\bar Y}_x^2, $$
which is explained variance over total variance or
squared multiple correlation
in Analysis of Variance terminology [7]. If $Y$ and $X$ are independent,
the distribution of $I$ is approximately that of a mixture of chi-square
variables with 1 degree of freedom $(df)$. Values of $I$ substantialy
greater than 1 are indications of some dependence.
The distribution of $J$ is
approximately that of chi-square with $k-1$ degrees of freedom if $X$ has
$k$ possible values with a substantial number of expected observations
for each $x$. In our applications this last condition rarely applies.

\subsection{Upper Bound  on $\theta_e$ Using $\theta_I$}
Since
\begin{equation}
\sum_i a_i^2 \leq (\sum_i |a_i|)^2 ,
\end{equation}
$\theta_I$ provides an
upper bound for $\theta_e$. Let 
\begin{equation}
\sum_x(f_d(x)-f_h(x))^2=\alpha(\sum_x|f_d(x)-f_h(x)|)^2,
\end{equation}
where $\alpha \leq 1$. Then
\begin{equation}
\theta_e=0.5-\sqrt{\theta_I/8\alpha},
\end{equation}
which may also be written as $\theta_I=8\alpha(.5-\theta_e)^2$.
Since $\alpha\leq 1$, $\theta_e \leq 0.5 - \sqrt{\theta_I/8}$.

This bound is rather weak since the ratio, $\sum_ia_i^2/(\sum_i|a_i|)^2$
can be anywhere between 1 and $1/k$ where $k$ is the number of terms
in the sum. In the appendix we prove that when $\sum_ia_i=0$, as is the 
case here of the difference of two probability distributions, the ratio
is no larger than 1/2. Thus we can use $\alpha \leq 0.5$ to derive the 
upper bound
\begin{equation}
\theta_e \leq 0.5-\sqrt{\theta_I/4}
\end{equation}
and 
\begin{equation}
\theta_I \leq 4(0.5-\theta_e)^2.
\end{equation}

\subsection{Estimators and Biases}
It remains for us to see how well the  naive estimators of these
parameters behave. The variances of these estimates depend mainly on
the second and fourth moments of $|n_{dx}-n_{hx}|$ .
 Let us first consider $E(I)$. Since $n_{dx}-n_{hx}$ is the difference
of two independent binomially distributed variables, the expectation of 
the sum of the squares is easily calculated.
\begin{align}
E(I)&=0.5n\sum_x(f_d(x)-f_h(x))^2 +  
  .5\sum_xf_d(x)(1-f_d(x))+f_h(x)(1-f_h(x)). \nonumber \\
&=n\theta_I + 1 - .5\sum_x(f_d(x)^2+f_h(x)^2). 
\end{align}
Then $I/n$ is of order 1 and the bias, is less than $1/n$. 

To calculate the expectation of
\begin{equation}
{\hat \theta}_e=0.5\sum_x \min(n_{dx},n_{hx})/n,
\end{equation}
the training estimator of $\theta_e$, we need
only calculate the contribution of each value of $x$. 
In general if $Z$ and $W$ are independent binomials with
parameters $n, p_Z $ and $p_W$  where $p_Z/p_W=r>1$
and $\lambda=np_Z$, we are interested in
the negative relative bias for each value of $x$.
\begin{equation}
b(n,\lambda,r)=\frac{n\,p_W-E(\min(Z,W))}{n\,p_W}.
\end{equation}
which can be computed directly. 
Since we expect the bias to be negative
$b$ should be positive. Table 1 shows that $b(n,\lambda,r)$ is almost
independent of $n$ for $n>100$. Figure 1 presents $b$ for $n=500$ where
the curves represent distinct values of $r$ and $\lambda$ is represented
by $la$.

\begin{table}[hp]
\caption{Negative Relative Bias for Training Estimate.}
\begin{center}
\begin{tabular}{c|ccccc}
                \multicolumn{6}{c}{$b(n,\lambda,r)$:  $\lambda=n*p_Z$; $r=p_Z/p_W$} \\ 
                 \multicolumn{6}{c}{ } \\
                                \multicolumn{6}{c}{$n=100$} \\ \hline
                                &\multicolumn{5}{c}{$r$} \\ \cline{2-6}
                 $\lambda$ & 40  &    5  &  1.25 &  1.062 &  1.016 \\ \hline
               40 &     0.0000 & 0.0000 & 0.0001 & 0.0123 & 0.0459 \\ 
               10 &     0.0001 & 0.0005 & 0.0283 & 0.0995 & 0.1469 \\
               2.5 &    0.0859 & 0.1052 & 0.2131 & 0.2929 & 0.3288 \\
               0.625 &  0.5368 & 0.5445 & 0.5827 & 0.6087 & 0.6202 \\
               0.156 & 0.8555 & 0.8563 & 0.8603 & 0.8632 & 0.8646 \\
               0.039 & 0.9617 & 0.9618 & 0.9620 & 0.9623 & 0.9624 \\ \hline
                \multicolumn{6}{c}{ } \\
                                \multicolumn{6}{c}{$n=500$} \\ \hline
                                &\multicolumn{5}{c}{$r$} \\ \cline{2-6}
                 $\lambda$ & 40  &    5  &  1.25 &  1.062 &  1.016 \\ \hline
               40  &    0.0000 & 0.0000 & 0.0004 & 0.0220 & 0.0617 \\
               10  &    0.0001 & 0.0007 & 0.0320 & 0.1064 & 0.1542 \\
               2.5  &   0.0880 & 0.1076 & 0.2164 & 0.2963 & 0.3322 \\
               0.625 &  0.5377 & 0.5453 & 0.5837 & 0.6098 & 0.6213 \\
               0.156 &  0.8556 & 0.8564 & 0.8604 & 0.8634 & 0.8647 \\
               0.039 &  0.9617 & 0.9618 & 0.9621 & 0.9623 & 0.9624 \\ \hline
                \multicolumn{6}{c}{ } \\
                                \multicolumn{6}{c}{$n=2,500$} \\ \hline
                                &\multicolumn{5}{c}{$r$} \\ \cline{2-6}
                 $\lambda$ & 40  &    5  &  1.25 &  1.062 &  1.016 \\ \hline
               40  &    0.0000 & 0.0000 & 0.0005 & 0.0239 & 0.0646 \\
               10  &    0.0001 & 0.0007 & 0.0327 & 0.1077 & 0.1556 \\
               2.5  &   0.0884 & 0.1081 & 0.2170 & 0.2970 & 0.3328 \\
               0.625 &  0.5378 & 0.5455 & 0.5839 & 0.6100 & 0.6215 \\
               0.156  & 0.8556 & 0.8564 & 0.8604 & 0.8634 & 0.8647 \\
               0.039  &  0.9617 & 0.9618 & 0.9621 & 0.9623 & 0.9624 \\ \hline
\end{tabular}
\end{center}
\label{tab:1}
\end{table}%

Figure 1 clearly shows that $b$ is close to 1 for small $\lambda$ and
close to 0 for large $\lambda$. This implies that the large
values of $n_{dx}$ and $n_{hx}$ contribute to reliable estimates, but the
values of $x$ with small entries are not helpful for the training estimator.

\begin{figure}[h]
\begin{center}
\includegraphics[width=0.8\textwidth]{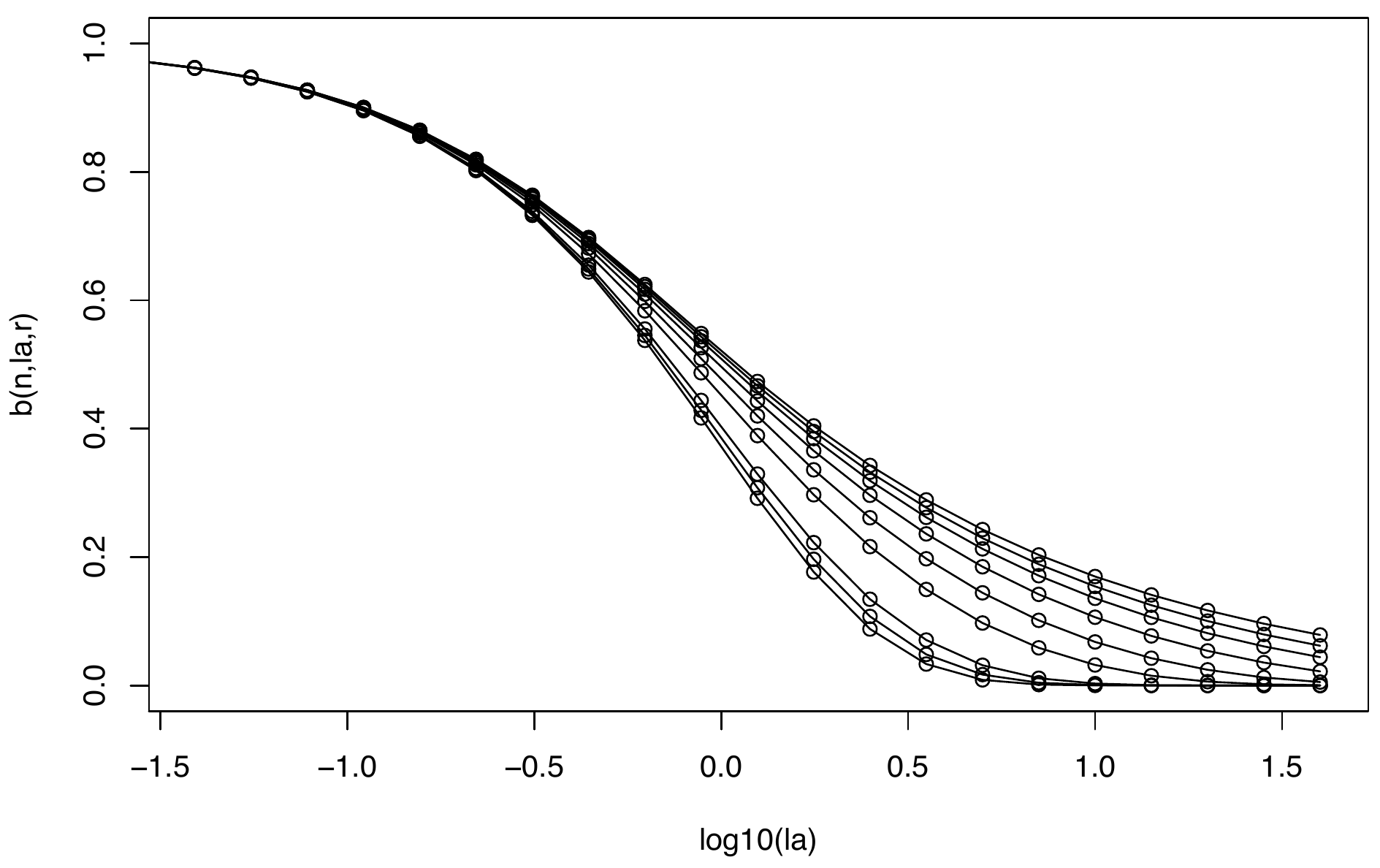}
\caption{Negative bias proportion versus $\log_{10}(\lambda)$ (labelled as ``log10(la)'') for 
$r = 1.1016$, $1.062$, $1.25$, $5$, and $40$. $n=500$. }
\end{center}
\end{figure}

The bias is
\begin{equation}
B_e=-0.5\sum_x(\min(f_d(x),f_h(x))b(n,\lambda(x),r(x)),
\end{equation} 
where $r(x)=\max(f_d(x),f_h(x))/\min(f_d(x),f_h(x))$ and where
$\lambda(x)=n\max(f_d(x),f_h(x))$.

This result suggests a correction for bias. We introduce
\begin{equation}
{\hat \theta}_{e1}=0.5\sum_{x:\min(n_{dx},n_{hx})>0} \frac{\min(n_{dx},n_{hx})}{n} (1-b(n,\lambda(x),r(x))).
\end{equation}				   

Both ${\hat \theta}_e$ and ${\hat \theta}_{e1}$
neglect the effect of those values of $x$  for which
there are 0 observations. We conjecture that it may be possible to use
 the Good-Turing [8] approach to compensate for this problem. A brief
description of this approach, used in code breaking, involves the following
problem. Suppose
that there are many species of fish in a lake. A random sample of n fish
are caught and we assume all fish in the lake are equally likely to  be
caught. What proportion of the fish in the lake are of species for which
none have been caught?. This {\it coverage} is estimated by $1/n$
times the number of {\it singletons} or the species for which only one
fish was caught.

Another estimator is called the {\it Out of sample estimator}
$\hat{\theta}_{eo}$. Unlike the other estimators which are based on the
data, this is a sort of hybrid in that it involves the underlying 
distributions which are unknown by the statistician, but known by the
simulator. It could also be considered a way to evaluate a method by using 
it many times (without adjusting on the basis of subsequent results).
It can be estimated by use of an independent sample. The
simulator who knows the model can evaluate the method by simulation or
by analysis. In 
our case the natural method consists of deciding $h$ for all future $x$
for which $n_{dx}<n_{hx}$ in this particular sample, deciding $d$ when
the inequality is reversed and choosing with probability 1/2 when the
two frequencies match. For the simulator who knows the model, 
this estimator is
\begin{align}
{\hat \theta}_o = & 0.5\left(\sum_xf_d(x)[1(n_{dx}<n_{hx})+1(n_{dx}=n_{hx})/2)] \right.\nonumber \\
                    & \left.+  \sum_xf_h(x)[1(n_{dx}>n_{hx})+1(n_{dx}=n_{hx})/2]\right),
\end{align}		      
where $1(A)$ is the characteristic function of $A$, {\it i.e.} 1 if 
$A$ is true and 0 if $A$ is false.
The contribution of $x$ for which $f_d(x)>f_h(x)$ 
to the expectation of this out of sample estimator is
$$0.5(f_d(x)(Pr(n_{hx}>n_{dx})+0.5Pr(n_{hx}=n_{dx})) 
      +f_h(x)(Pr(n_{hx}<n_{dx})+0.5Pr(n_{hx}=n_{dx}))). $$
In terms of $Z$ and $W$ above, let 
$$a(x)=a(n,\lambda(x),r(x))=Pr(Z<W)+0.5Pr(Z=W). $$
 Thus the contribution of $x$
to the estimate of the error
probability is $(f_d(x)a(x)+f_h(x)(1-a(x)))/2$ to be compared with 
$f_h(x)/2$, the contribution to $\theta_e$.

A similar result applies to those
values of $x$ for which $f_d(x)<f_h(x)$ and for which $f_d(x)=f_h(x)$.
This  gives rise to a relative contribution to bias of 
$$b_o(x)=(r(x)-1)a(n,\lambda(x),r(x)). $$ 

Table 2 indicates that $a$, and therefore $b_o$, are almost independent
of $n$ for $n>100$, given $\lambda$ and $r$. Figure 2 plots $a$ for $n=500$,
where the curves correspond to distinct values of $r$.

\begin{table}[htp]
\caption{$a(n,\lambda,r)=\mbox{Pr}(Z<W)+0.5\mbox{Pr}(Z=W)$, 
                             $\lambda=np_Z$; $r=p_Z/p_W$.}
\begin{center}
\begin{tabular}{c|ccccc}
                                  \multicolumn{6}{c}{$n=100$} \\ \hline
                                &\multicolumn{5}{c}{$r$} \\ \cline{2-6}
                 $\lambda$ & 40  &    5  &  1.25 &  1.062 &  1.016 \\ \hline
                 40     & 0.0000 & 0.0000 & 0.1192 & 0.3675 & 0.4638 \\
                10       & 0.0001 & 0.0059 & 0.3115 & 0.4449 & 0.4852 \\
                2.5     & 0.0488 & 0.1201 & 0.4083 & 0.4740 & 0.4930 \\
                0.625   & 0.2739 & 0.3199 & 0.4593 & 0.4884 & 0.4969 \\
                0.156   & 0.4295 & 0.4429 & 0.4863 & 0.4960 & 0.4989 \\
                0.039   & 0.4813 & 0.4847 & 0.4962 & 0.4989 & 0.4997  \\ \hline
                 \multicolumn{6}{c}{ } \\
                               \multicolumn{6}{c}{$n=500$} \\ \hline
                                &\multicolumn{5}{c}{$r$} \\ \cline{2-6}
                 $\lambda$ & 40  &    5  &  1.25 &  1.062 &  1.016 \\ \hline
                40      & 0.0000 & 0.0000 & 0.1639 & 0.3915 & 0.4707 \\
                10      & 0.0001 & 0.0070 & 0.3181 & 0.4471 & 0.4858 \\
                2.5     & 0.0499 & 0.1217 & 0.4092 & 0.4742 & 0.4931 \\
                0.625   & 0.2743 & 0.3202 & 0.4595 & 0.4885 & 0.4969 \\
                0.156   & 0.4296 & 0.4429 & 0.4863 & 0.4960 & 0.4989 \\
                0.039   & 0.4813 & 0.4847 & 0.4962 & 0.4989 & 0.4997 \\ \hline
                 \multicolumn{6}{c}{ } \\
                               \multicolumn{6}{c}{$n=2,500$} \\  \hline
                                &\multicolumn{5}{c}{$r$} \\ \cline{2-6}
                 $\lambda$ & 40  &    5  &  1.25 &  1.062 &  1.016 \\ \hline
                40      & 0.0000 & 0.0000 & 0.1712 & 0.3949 & 0.4716 \\
                10      & 0.0001 & 0.0073 & 0.3193 & 0.4475 & 0.4860 \\
                2.5     & 0.0501 & 0.1220 & 0.4093 & 0.4743 & 0.4931 \\
                0.625   & 0.2744 & 0.3203 & 0.4595 & 0.4885 & 0.4969 \\
                0.156   & 0.4296 & 0.4429 & 0.4863 & 0.4960 & 0.4989 \\
                0.039   & 0.4813 & 0.4847 & 0.4962 & 0.4989 & 0.4997 \\ \hline
\end{tabular}
\end{center}
\label{tab:2}
\end{table}%

\begin{figure}[h]
\begin{center}
\includegraphics[width=0.8\textwidth]{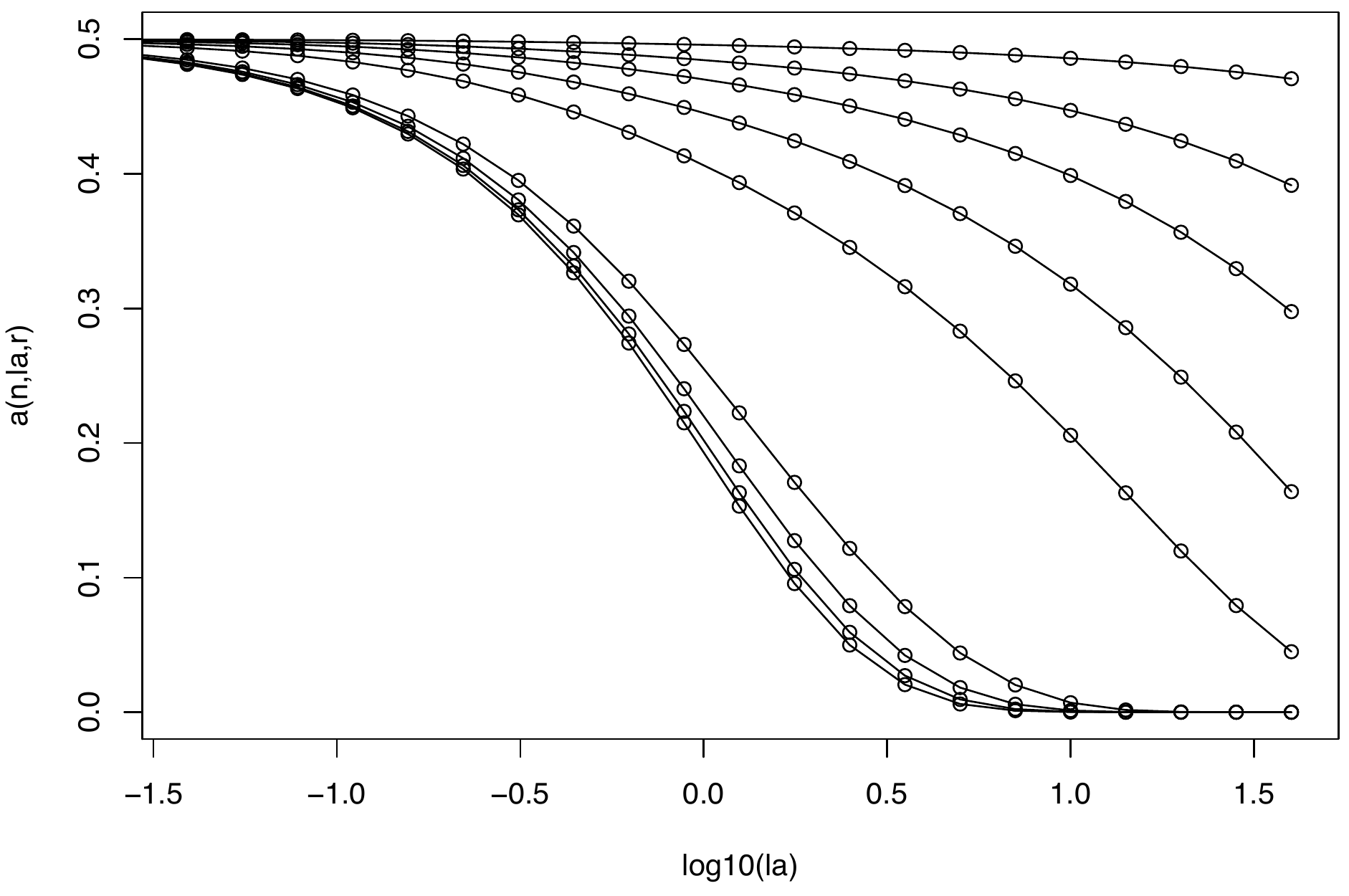}
\caption{$a(n, \lambda, r)=\mbox{Pr}(Z<W)+0.5\mbox{Pr}(Z=W)$ versus $\log_{10} (\lambda)$ for
$r = 1.1016$, $1.062$, $1.25$, $5$, and $40$. $n=500$.  }
\end{center}
\end{figure}

The bias of the out  of sample  estimator can be expressed as
\begin{equation}
B_{eo}= 0.5\sum_x\min(f_d(x),f_h(x))b_o(x),
\end{equation}   
and a modified estimate would be 
\begin{equation}
{\hat \theta}_{eo1} = {\hat \theta}_o - 0.5\sum_x\min(f_d(x),f_h(x))b_o(x).
\end{equation}

The out of sample estimator tends to overestimate $\theta_e$. For some
readers there may be an apparent paradox when these two estimators,
using the same method, bias the result in different directions. However
the training estimate uses the method on the data set from which the
method was derived, while the out of sample estimate is evaluated on how
well it will do on all future data. There, the method is suboptimal, since
it does not use the actual probabilities $f_d$ and $f_h$, but estimates of
these probabilities.  

\section{DESCRIPTION OF PARTITION RETENTION}
We present here a very concise description of the major idea of 
the partition retention method for $VS$. Given a small group 
of discrete explanatory variables,
we evaluate $I$ for this group and for each subset where one of the
group is dropped. If none of these subsets lead to an increase in $I$,
the entire group is {\itshape retained}. Otherwise that element, 
the dropping of which leads to 
the largest increase in the value of $I$, is eliminated from
the group. This reduction  procedure
is then applied to the remaining group and repeated until we reach
the subgroup where none are dropped. This backward 
procedure is repeated and applied
to many groups selected at random. Then those variables that are 
retained very often are candidates to be taken seriously.

For large $m$, the variables that show up very well at first
in this backward selection approach, are used to select new 
subsets which may succeed in {\itshape resuscitating}
influential variables that did not show up well at first, but interact
with some that did. This is done by forming subsets which contain some
of the {\it good} variables and some of the others.
For very  large $m$ computational limits may force us to go through
several stages
starting with groups consisting of only  one or two variables.

This backward selection method within each stage 
distinguishes the method from some competitors
such as Random Forests [9],
where there is a forward selection approach. We
believe that choices of first optimality leads to less reliable results
than ours of first discarding the least valuable.
\section{GROUPS OF SNPS}
It should be noted that
in the successful use of $PR$ referred to previously, the investigators 
found a substantial number of small interacting subgroups, each of which
was capable of rather weak predictivity, but acting together provided
a major improvement. Thus it is of value to be able to find small
subgroups which can be the basis for average error rates close to,
but less than 1/2. In the following we shall study how $\theta_e$ and $\theta_I$ are related for small groups of SNPs, concentrating at first on the case where only one of the group is influential. 

\subsection{Formal Notation}
Let us specialize to the case where we observe a few independent SNPs, with common minor allele frequency (MAF) $p$,  
only the first of which is influential.
We note that when $m$ is large it is unlikely that a group of 6 snps
selected at random will have more than one of several interacting
influential snps among them. With groups of 6 snps, each of which can
have 3 possible values we deal with with $279=3^6$ possible  values of
$x$. The use of substantially larger groups will do little to increase the 
possibility of finding two interacting snps in the same group, but will
make it likely that for almost all of the values of $x$\, 
$n_x$ will be 0 or 1 
more or less at random, and contribute little useful information.
Although this case of only
one influential snp in the group involves no interactions, 
it is useful to help us understand why $I$ works well.

Let $X = (U,V)$ where $U$ represents the first snp which can assume the values
0,1 and 2. The minor allele frequency $(MAF)$ for this snp is $p$ and the
above values of  $U$ are assumed with probabilities 
$f_U(u)=(1-p)^2,2p(1-p),p^2$
for  $u=(0,1,2)$. Treating $f$ as the discrete probability density, we
assume that $f_{Y|X}(h|x)=t(u)$ where $t$ is a decreasing
function of $u$.

Some elementary calculations yield
$$ f_X(x)=f_U(u)f_V(v) $$
assuming $U$ and $V$ are independent.
\begin{align*}
f_{Y|X}(d|x)&=1-t(u) \\
f_{XY}(x,h) &=t(u)f_U(u)f_V(v)\\
f_{XY}(x,d) & =(1-t(u))f_U(u)f_V(v)\\
f_Y(h)&=\sum_xf_{XY}(x,h)=\sum_ut(u)f_U(u)\\
f_{X|Y}(x|h)&=t(u)f_U(u)f_V(v)/f_Y(h) \\
f_{X|Y}(x|d)&=(1-t(u))f_U(u)f_V(v)/f_Y(d) 
\end{align*}
where $f_Y(d)=1-f_Y(h)$. Note that what we previously referred to as
$f_d(x)$ and $f_h(x)$ in the case control example are presented here 
as $f_{X|Y}(x|d)$ and $f_{X|Y}(x|h).$

To get a concrete feeling for the conditional distributions, it may
help to refer to a specific (artificial) example. In the
following example with the group of 6 SNPs and 279 possible values of $X$,
each with small probability, let the $MAF$ of each of the SNPs be 0.2, and
let $t(u)$ take on the values $(0.97,0.60,0.40)$ for $u=(0,1,2)$. The
SNP $U$ takes on these three values with probabilities 
$f_U(u)$ given by
$(0.64,0.32,0.04)$. Then we can calculate $f_Y(d)=0.171,
f_Y(h)=0.829, f_{U|Y}(u|d)=(0.112,0.748,0.140)$ and
$f_{U|Y}(u|h)=(0.749,0.232,0.019)$
 The ratios of these conditional probabilities are $(6.688,0.310,0.138)$.

Note that $f_{X|Y}=f_{U|Y}f_V$ and the likelihood ratios for the conditional
distributions of $X$ are limited to the same three possible values
depending on the value of $U$. Although
the values of $f_{U|Y}$ are substantial, almost all 729 possible values of
$X$ come with  small probabilities. Note also that the likelihood
ratio for $u=0$ is greater than 1 and the other values are 
considerably less than 1.

\subsection{Calculations}
Now let us calculate $\theta_e$,  and $\theta_I$.
First, since $\sum_vf_V(v)=1$,
\begin{equation}
\theta_e=.5(f_{U|Y}(0|d)+f_{U|Y}(1|h)+f_{U|Y}(2|h)).
\end{equation}

In our special case $\theta_e = 0.182$. and is not affected by the  5 
non-influential variables. This is the best we can hope
to do in predicting a random new subject.
Moreover, the statistician who does not know the 
underlying probabilities may not be able to do so well. While we can
ignore the five useless variables in the group, the statistician
using this group may find the presence of the other variables distracting
and leading him to have poorer performance.

Now we calculate 
\begin{equation}
\theta_I=.5\sum_u(f_{U|Y}(u|d)-f_{U|Y}(u|h))^2\sum_vf_V(v)^2.
\end{equation}

In our special case where $\sum_vf_V(v)^2 = .03574$ we obtain 
$\theta_I=0.0123$. Ignoring the term $\sum_vf_V(v)^2$ , that is  the
the factor by which  $\theta_I$ tends to be degraded by the presence of the
noninformative variables in the group,
we introduce the parameter
\begin{equation}
\theta_{I_0} = \theta_I/\sum_vf_V(v)^2
\end{equation}
to be compared with $\theta_e$. In our special case $\theta_{I_0}=0.344$
and the corresponding upper bound on $\theta_e$ is 0.207.
Simple graphs indicate that, as $t(0)$ decreases and $t(1)$
and $t(2)$ increase,
$\theta_{I_0}$ decreases and $\theta_e$ increases, indicating 
that these parameters become  less favorable for prediction. 
Table 3 lists several functions $t(u)$ and Figure 3 presents ($\theta_e$, $\theta_{I_0}$) along curves representing a $t(u)$ as the $MAF=p$ varies from $0.001$ to $0.3$. Figure 3 shows that the two parameters $\theta_e$ and $\theta_{I_0}$ are closely correlated. 
The upper bound that $\theta_{I_0}$
imposes on $\theta_e$ is represented in Figure 3 by the dashed curve.
%
Table 3 presents $\theta_e$ and $\theta_{I_0}$ for various values of $p$ and $t$.

\begin{figure}[ht]
\begin{center}
\includegraphics[width=0.8\textwidth]{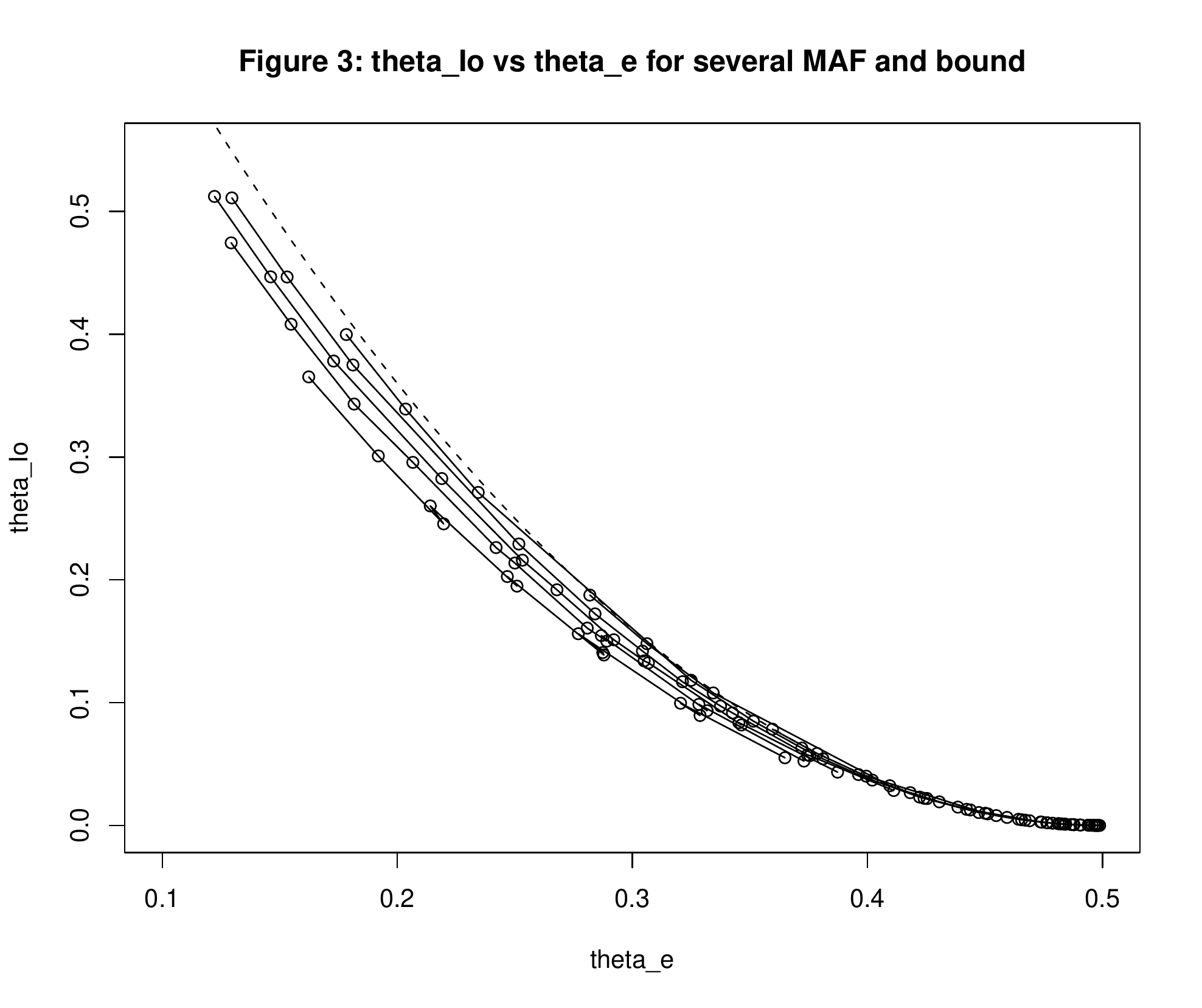}
\caption{$\theta_{I_0}$ versus $\theta_{e}$ for several MAF and bound. }
\end{center}
\end{figure}

\begin{table}[htp]
\caption{Various choices of $t(u) = P(h | u)$  for a single influential variable.}
\begin{center}
\begin{tabular}{cc|ccc} \hline
                 &	& \multicolumn{3}{c}{$u$} \\ \cline{3-5}
                 & &   0  &     1   &   2  \\ \hline
                 &  $t_1$   & 0.97   & 0.4    & 0.2  \\
                 &  $t_2$   & 0.97   & 0.5    & 0.3 \\
                 &  $t_3$   & 0.97   & 0.6    & 0.4 \\
                 &  $t_4$   & 0.90   & 0.4    & 0.2 \\
                 &  $t_5$   & 0.90   & 0.5    & 0.3 \\
               Choices of &  $t_6$   & 0.85   & 0.4    & 0.2 \\
              $t(u) = P(h | u)$  &  $t_7$   & 0.90   & 0.6    & 0.4 \\
               &  $t_8$   & 0.85   & 0.5    & 0.3  \\
                 &  $t_9$   & 0.80   & 0.4    & 0.2 \\
                 & $t_{10}$   & 0.85   & 0.6    & 0.4 \\
                 &  $t_{11}$   & 0.80   & 0.5    & 0.3 \\
                 &  $t_{12}$   & 0.80   & 0.6    & 0.4    \\   \hline
\end{tabular}
\end{center}
\label{tab:3a}
\end{table}%

As  $p$,  increases the curves of $\theta_{I_0}$
 vs $\theta_e$  move to the left, indicating improved predictability, until
 $p$ reaches about 0.15. For  some $t$ functions $\theta_e$ reaches a 
 minimum and starts to increase for $p>0.15$.
 Sometimes, for $p$ greater than 0.15,  we have small regions where $I_0$ 
changes slowly in the same direction as $\theta_e$. In these regions 
small changes in $I_0$ are  unreliable indicators of
corresponding changes in $\theta_e$.

The quantity $\sum_vf_V(v)^2$ is a product of factors for each of the
SNP's of $V$. For a MAF of $p$ the factor contributed by the SNP is
$p^4+(2p(1-p))^2+(1-p)^4$. This function is symmetric about $p=0.5$ and
decreases rapidly from 1 at $p=0$ to 0.5136 at $p=0.2$ and more slowly
to 3/8 at $p=0.5$.
In our case where each $p=0.2$ we have
$\sum_vf_V(v)^2=(0.5136)^5=0.0357$.

We digress momentarily to emphasize that while $\theta_e$ does not change
by adjoining irrelevant variables to the group, $\theta_I$ is degraded.
This fact supports the strategy of Partition Retention that consists of
discarding each variable that diminishes $I$.

In Table 4  we present the results of a set of simulations designed to check on our estimates of the biases of the training and out of sample estimates and the corrections for these biases.  The inputs are
the MAF and $t$ functions and $n$ and $m$, the number of repetitions.
For each case we calculate $\theta_e, \theta_{I_0}, B, B_o$ and the bound on
$\theta_e$. In addition we calculate the average and standard deviations
based on the $m$ repetitions of the observed biases of the training estimate,
the correction for the training estimate, and the out of sample estimate.
The averages of the biases for the training estimate and the out of sample
estimate are reasonably close to their expectations. 

Three comments are worth making. The correction for the training estimate
seems to reduce the bias by a factor varying from 0.5 to 0.1. 
The upper bound on $\theta_e$ based on $\theta_{I_0}$ 
is generally surprisingly tight. This calculation provides a hint
that the estimates generally have moderate sampling variability.

\begin{table}[htp]
\caption{Biases for groups with 1 influential variable.}
\begin{center}
\begin{tabular}{cccccc|ccc|cc}           \hline
                       \multicolumn{6}{c}{}  &   \multicolumn{3}{|c|}{$t(u)$}    &  \\ \cline{7-9}
                                      \multicolumn{6}{c|}{MAF}&0&    1&   2&  $m$& $n$  \\ \hline
             0.2 & 0.2 & 0.2 & 0.2 & 0.2 & 0.2   & 0.97 & 0.6 & 0.4   & 25  & 100 \\
             0.2 & 0.2 & 0.2 & 0.2 & 0.2 & 0.2   & 0.70 & 0.6 & 0.5   & 25  & 200 \\
             0.2 & 0.2 & 0.2 & 0.2 & 0.2 & 0.2   & 0.80 & 0.5 & 0.2   & 25  & 100 \\
             0.1 & 0.1 & 0.1 & 0.2 & 0.0 & 0.0   & 0.97 & 0.6 & 0.4   & 25  & 200 \\
             0.1 & 0.1 & 0.1 & 0.2 & 0.0 & 0.0   & 0.70 & 0.6 & 0.5   & 25  & 100 \\
             0.1 & 0.1 & 0.1 & 0.2 & 0.0 & 0.0   & 0.80 & 0.5 & 0.2   & 25  & 200 \\
             0.2 & 0.1 & 0.1 & 0.1 & 0.2 & 0.0   & 0.97 & 0.6 & 0.4   & 25  & 100 \\
             0.2 & 0.1 & 0.1 & 0.1 & 0.2 & 0.0   & 0.70 & 0.6 & 0.5   & 25  & 200 \\
             0.2 & 0.1 & 0.1 & 0.1 & 0.2 & 0.0   & 0.80 & 0.5 & 0.2   & 25  & 200 \\
		\multicolumn{11}{c}{} \\
		
\end{tabular}

\begin{tabular}{ccccccccccc}           
$\theta_e$ &  $b$ &    $b_o$  &     $\bar{b}$ &  $s_b$  &    $\bar{b}_1$ &  $s_{b_1}$ &  $\bar{b}_o$ &  $s_{b_o}$ & $\theta_{lo}$ &  bound \\ \hline
0.182  & 0.058 & 0.132& -0.057 & 0.024 &-0.011 & 0.033 & 0.145 & 0.016 & 0.343 & 0.207 \\
0.443 & 0.159 & 0.047& -0.157 & 0.018  &-0.075 & 0.031 & 0.051 & 0.004 & 0.011 & 0.449 \\
0.324 & 0.124 & 0.113 & -0.124 & 0.019 &-0.048 & 0.027 & 0.128 & 0.021 & 0.101 & 0.341 \\
0.181 & 0.005 & 0.015 & -0.006 & 0.017  & 0.001 & 0.018 & 0.017 & 0.005 & 0.375 & 0.194 \\
0.463 & 0.074 & 0.026 & -0.072 & 0.017  &-0.028 & 0.028 & 0.028 & 0.008 & 0.005 & 0.465 \\
0.374 & 0.020 & 0.024 & -0.017 & 0.018  & 0.002 & 0.020 & 0.023 & 0.012 & 0.058 & 0.380 \\
0.182 & 0.020 & 0.051 & -0.025 & 0.022  &-0.004 & 0.024 & 0.059 & 0.014 & 0.343 & 0.207 \\
0.443 & 0.070 & 0.034 & -0.076 & 0.018  &-0.035 & 0.026 & 0.037 & 0.013 & 0.011 & 0.449 \\
0.324 & 0.029 & 0.037 & -0.026 & 0.017  & 0.001 & 0.021 & 0.045 & 0.012 & 0.101 & 0.341 \\
\end{tabular}
\end{center}

\begin{itemize}
\item $b=$ negative bias of training estimate;
\item $b_o=$ bias for out of sample estimate;
\item $\bar{b}=$ average bias for training estimate in 25 simulations;
\item $s_b=$ standard deviation of bias for training estimate;
\item $\bar{b}_1=$ average bias of adjusted estimate;
\item $s_{b_1}=$ standard deviation of adjusted bias;
\item $\bar{b}_o=$ average out of sample estimate;
\item $s_{b_o}=$ standard deviation of out of sample estimate;
\item bound$=0.5-\sqrt{\theta_{Io}/4}$ on $\theta_e$.
\end{itemize}

\label{tab-4}
\end{table}%

\subsection{Two Influential Variables}
The main point of the $PR$ method was to take advantage of the possible
interactions among influential variables which may not indicate much
marginal effect by themselves. Thus
we now consider the case where there are two,
possibly interacting influential variables
in the group under consideration. For moderately  large $m$, we will find
such groups by using a large number of randomly  selected groups. For larger
$m$ Partition Retention may require {\it resuscitation} and for very large
$m$ we may require several stages, starting with one variable at a time,  and
moving to two variables and then more.

In this  case we can write
$X=(U,V)$ where $U=(U_1,U_2)$ are the influential SNPs
and $V$ represents the other variables  
in the group and which are assumed to be independent of $U$
and each other. The only change that takes place is that the function
$t$ now takes on 9 possible values corresponding to the possible
values of $U$. Typically we would expect $t$ to decline as u varies
over ${(00),(10),(01),(20),(02),(11),(21),(12),(22)}$
although that is not necessarily the case.

As a special case we take the function $t(u)$ which assumes the values
($0.95$, $0.75$, $0.7$, $0.60$, $0.50$, $0.20$, $0.15$, $0.10$, $0.05$) for the values
of $u$ listed above. As before we take a group of 6 SNP's with MAF 0.2
and we may calculate 
$$\theta_e=0.5\sum_u \min(f_{Y|U}(d|u),f_{Y|U}(h|u))=0.269$$
which is not affected by the 4 extraneous variables. We can compute
$\theta_{I_0}=0.1447$ using the same formula as before and $\theta_I=
\theta_{I_0}\sum_vf_V(v)^2 = 0.1447*0.06958= 0.01001$.
Since there are only 4 variables
among the irrelevant ones in the group, $\sum f_V(v)^2=0.06958$ in our
special case. 

\begin{figure}[h]
\begin{center}
\includegraphics[width=0.8\textwidth]{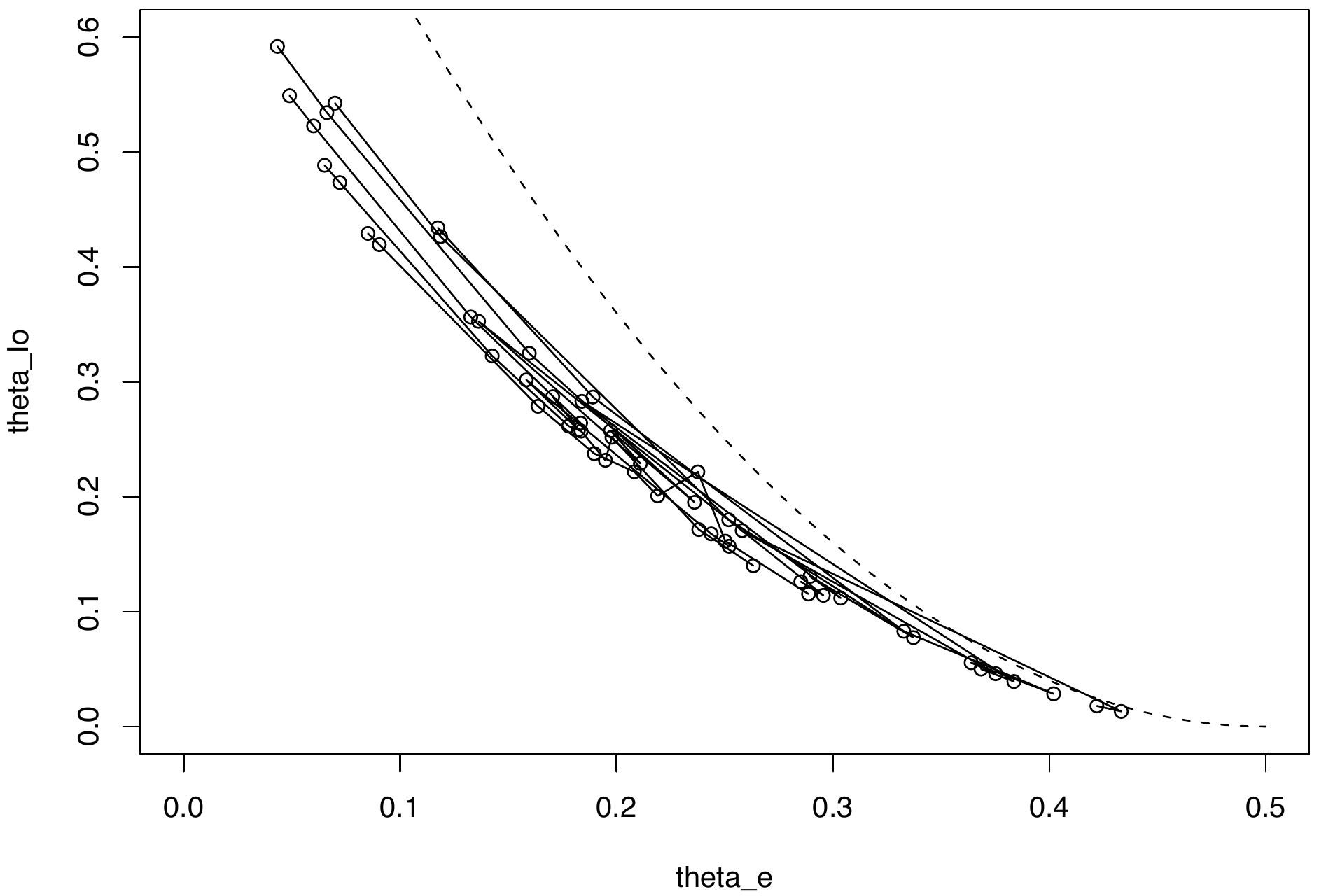}
\caption{$\theta_{I_0}$ versus $\theta_e$ for several MAF and bound; two influential variables.}
\end{center}
\end{figure}

In Figure 4 we present the graph of $\theta_{I_0}$ vs. $\theta_e$ for
6 values of MAF and  9 versions of $t(u)=f_{Y|U}(h|u)$ tabulated in
Table 5.  Each row of $t(u)$ represents a distinct curve. It is clear
from the graph that $\theta_{I_0}$ is closely correlated with 
$\theta_e$. The bound on $\theta_e$ provided by $\theta_{I_0}$ is 
represented by the dashed curve. This bound seems to be relatively
strong when $\theta_e$ is close to 0.5, {\it i.e.} when the error
rate is high.

\begin{table}[ht]
\caption{Various choices of $t(u)=P(h|u)$ for the case of two influential SNPs in a group of 6 with a common MAF of $p$.}
\begin{center}
\begin{tabular}{cc|ccccccccc} \hline
                  && \multicolumn{9}{c}{$u$} \\ \cline{3-11}
        &  & 00 & 10 &  01  & 20  &  02  &  11  &   21  &   12  &   22 \\ \hline  
         &$t_1$  & 0.99 & 0.20 & 0.15 & 0.10 & 0.08 & 0.02 & 0.010 & 0.004 & 0.001 \\
         & $t_2$  & 0.98 & 0.20 & 0.15 & 0.15 & 0.10 & 0.01 & 0.005 & 0.005 & 0.001 \\
         & $t_3$  & 0.95 & 0.40 & 0.30 & 0.20 & 0.10 & 0.02 & 0.010 & 0.005 & 0.002 \\
      Choices of   & $t_4$  & 0.90 & 0.40 & 0.30 & 0.20 & 0.10 & 0.02 & 0.010 & 0.005 & 0.002 \\
      $t(u) = P(h|u)$  & $t_5$  & 0.95 & 0.60 & 0.40 & 0.40 & 0.20 & 0.02 & 0.010 & 0.005 & 0.001 \\
        & $t_6$  & 0.95 & 0.60 & 0.50 & 0.30 & 0.10 & 0.05 & 0.020 & 0.015 & 0.010 \\
        & $t_7$  & 0.99 & 0.75 & 0.75 & 0.70 & 0.70 & 0.01 & 0.008 & 0.008 & 0.002 \\
       &  $t_8$  & 0.90 & 0.60 & 0.50 & 0.30 & 0.10 & 0.05 & 0.020 & 0.015 & 0.010 \\
       &  $t_9$  & 0.95 & 0.75 & 0.70 & 0.60 & 0.50 & 0.40 & 0.350 & 0.300 & 0.200 \\ \hline
 \end{tabular}

\end{center}
\label{tab-5}
\end{table}%

In Table 6 we present the input and output for some simulations involving
two influential interacting variables in a group of 6 SNPs. 
Table 6 is parallel to Table 4 and the comments following Table 4
apply here too. This should also be the case when we have more than 2 
influential variables in the group.

\begin{table}[hp]
\caption{Biases for groups with 2 influential variables}
\begin{center}
\begin{tabular}{cccccccccccccccccc} \hline          
    &&&&&&                              	\multicolumn{9}{c}{$t(u1,u2)$} & & \\  \cline{7-15}
 \multicolumn{6}{c}{MAF}  & 00 &  01 & 02 & 10 & 11 & 12 & 20 & 21 &  22 & $m$  &  $n$ \\ \hline
 0.2 & 0.2 & 0.2 & 0.2 & 0.2 & 0.2   & 0.97 & 0.6 & 0.4 & 0.7 & 0.6 & 0.4 & 0.4 & 0.6 & 0.6   & 25 & 100 \\
 0.2 & 0.2 & 0.2 & 0.2 & 0.2 & 0.2   & 0.70 & 0.6 & 0.5 & 0.6 & 0.6 & 0.4 & 0.4 & 0.6 & 0.6   & 25 & 200 \\
 0.2 & 0.2 & 0.2 & 0.2 & 0.2 & 0.2   & 0.80 & 0.5 & 0.2 & 0.7 & 0.6 & 0.4 & 0.4 & 0.6 & 0.6   & 25 & 100 \\
 0.1 & 0.1 & 0.1 & 0.2 & 0.0 & 0.0   & 0.97 & 0.6 & 0.4 & 0.6 & 0.6 & 0.4 & 0.4 & 0.6 & 0.6   & 25 & 200 \\
 0.1 & 0.1 & 0.1 & 0.2 & 0.0 & 0.0   & 0.70 & 0.6 & 0.5 & 0.5 & 0.5 & 0.4 & 0.4 & 0.5 & 0.5   & 25 & 100 \\
 0.1 & 0.1 & 0.1 & 0.2 & 0.0 & 0.0   & 0.80 & 0.5 & 0.2 & 0.5 & 0.5 & 0.4 & 0.4 & 0.5 & 0.5   & 25 & 200 \\
 0.2 & 0.1 & 0.1 & 0.1 & 0.2 & 0.0   & 0.97 & 0.6 & 0.4 & 0.5 & 0.4 & 0.3 & 0.3 & 0.4 & 0.4   & 25 & 100 \\
 0.2 & 0.1 & 0.1 & 0.1 & 0.2 & 0.0   & 0.70 & 0.6 & 0.5 & 0.5 & 0.4 & 0.3 & 0.3 & 0.4 & 0.4   & 25 & 200 \\
 0.2 & 0.1 & 0.1 & 0.1 & 0.2 & 0.0   & 0.80 & 0.5 & 0.2 & 0.5 & 0.4 & 0.3 & 0.3 & 0.4 & 0.4   & 25 & 200 \\
		\multicolumn{17}{c}{} \\	
\end{tabular}

\begin{tabular}{ccccccccccc}       
$\theta_e$ &  $b$ &    $b_o$  &     $\bar{b}$ &  $s_b$  &    $\bar{b}_1$ &  $s_{b_1}$ &  $\bar{b}_o$ &  $s_{b_o}$ & $\theta_{lo}$ &  bound \\ \hline
  0.264 & 0.111 & 0.108 & -0.107 & 0.026 & -0.041 & 0.035 & 0.136 & 0.012 & 0.137   & 0.315 \\
  0.439 & 0.157 & 0.046 & -0.160 & 0.014 & -0.081 & 0.024 & 0.055 & 0.009 & 0.009   & 0.453 \\
  0.359 & 0.152 & 0.090 & -0.154 & 0.027 & -0.074 & 0.043 & 0.107 & 0.017 & 0.046   & 0.393 \\
  0.182 & 0.006 & 0.016 & -0.007 & 0.018 & 0.002 & 0.020 & 0.027 & 0.008 & 0.271   & 0.240 \\
  0.421 & 0.050 & 0.036 & -0.051 & 0.031 & -0.013 & 0.041 & 0.045 & 0.027 & 0.017   & 0.434 \\
  0.336 & 0.013 & 0.021 & -0.012 & 0.021 & 0.004 & 0.023 & 0.030 & 0.012 & 0.072   & 0.366 \\
  0.200 & 0.027 & 0.055 & -0.028 & 0.021 & -0.000 & 0.027 & 0.080 & 0.017 & 0.226   & 0.262 \\
  0.413 & 0.057 & 0.036 & -0.063 & 0.016 & -0.025 & 0.022 & 0.042 & 0.013 & 0.019   & 0.431 \\ 
  0.317 & 0.029 & 0.038 & -0.031 & 0.024 & -0.003 & 0.029 & 0.057 & 0.023 & 0.085   & 0.355 \\ \hline
\end{tabular}
\end{center}
\begin{itemize}
\item $b=$ negative bias of training estimate;
\item $b_o=$ bias for out of sample estimate;
\item $\bar{b}=$ average bias for training estimate in 25 simulations;
\item $s_b=$ standard deviation of bias for training estimate;
\item $\bar{b}_1=$ average bias of adjusted estimate;
\item $s_{b_1}=$ standard deviation of adjusted bias;
\item $\bar{b}_o=$ average out of sample estimate;
\item $s_{b_o}=$ standard deviation of out of sample estimate;
\item bound $=0.5-\sqrt{\theta_{Io}/4}$ on $\theta_e$.
\end{itemize}
\label{tab-6}
\end{table}%

\section{A REAL DATA APPLICATION}

We have referred to the success [5] previously derived from the vant Veer breast
cancer data [9] where the standard error rate in prediction  of about 30
percent was reduced to 8\% using $PR$ and $I$ for the $VS$ part of the
analysis. The pattern recognition part used 18 small groups (modules) of
interacting variables, none of which had much predictive power by
themselves. In Table 7 we present the module of 5 variables which had
the highest $I$ score and the best prediction performance of these
modules, based on 
cross validation and an independent testing set in [9] . The error rate
using this module was estimated to be bounded by 41\%.  We see that the 
individual significance levels and combined significance level of these
5 variables were rather unimpressive, considering that we had available
4,918 candidates.

\begin{table}[h]
\caption{A predictive group}
\begin{center}
\begin{tabular}{llll}
     & Systematic name  & Gene name     &    Marginal p-value \\ \hline 
  1  & Contig45347\_RC    &  KIAA1683    &  0.008  \\  
  2  &  NM\_005145   &       GNG7        &     0.54  \\  
  3  &   Z34893     &     ICAP-1A      &      0.15 \\  
  4  &  NM\_006121  &        KRT1       &      0.9  \\  
  5  &  NM\_004701   &       CCNB2    &       0.003  \\  \hline
&&& \\
     &Joint I-score  &  Joint p-value  &   Family-wise threshold \\ 
     &    2.89             & 0.005                  &   0.0000007 \\
    &&& \\
    & $n = 97$ & \multicolumn{2}{l}{Bound on error prob: 0.414} \\
\end{tabular}
\end{center}
\label{tab-7}
\end{table}%

\section{SOME REMAINING ISSUES}
The variance of the estimates of $\theta_e$ and $\theta_I$
have been computed by formula, but are complicated.
Extensive simulations have been carried out  on models with several
influential variables and several noninfluential variables, and 
these variances are modest. It would be nice to  have simple bounds. 

We have assumed that $h$ and $d$ are equally  likely and the costs of
false positive and false negative were equal. Suppose that the two 
costs of error, when $h$ and $d$ are the states, are respectively
$c_h$ and $c_d$, and these two states have prior probabilities
$\pi_h$ and $\pi_d=1-\pi_h$. Then the optimal choice, for the decider
without data, is to 
select according to the smaller of $\pi_dc_d$ and $\pi_hc_h$ giving rise
to an expected cost equal to the lower of these amounts. 

Given the  data $x$, we replace the priors by the posterior probabilities
and we seclect according to the smaller of $\pi_dc_df_d(x)$ and 
$\pi_hc_hf_h(x)$ giving rise to an expected cost of
$$ \theta_C=\sum_x\min(\pi_dc_df_d(x),\pi_hc_hf_h(x)).$$
Since
$$\theta_C+\sum_x\max(\pi_dc_df_d(x),\pi_hc_hf_h(x))=C=\pi_dc_d+\pi_hc_h$$
we have
$$\sum_x|\pi_dc_df_d-\pi_hc_hf_h|=C-2\theta_C.$$
This result suggests that we use a modified version of $I$, depending on
$$\sum_x(\pi_dc_df_d(x)-\pi_hc_hf_h(x))^2,$$
for variable selection.

We have not addressed the problems of deciding among more than two
alternatives nor that of predicting a continuous dependent variable.
Generalizing to the case of a finite number of alternatives should not
be very difficult.

While we deal with small groups of explanatory variables, it would be
desirable to know how to extend our bound on the ideal error probability
when given a small group of modules, each consisting of a small group
of explanatory variables. A direct attack using our methods would fail
if the total number of variables among the modules is so large that
almost all $n_{dx}$ and $n_{hx}$ are small.

This
discussion omits the substantial problem of dealing with the pattern
recognition problem once a $VS$ choice has been made, and the problem
of dealing  with continuous explanatory variables.

\section{CONCLUSIONS}
If $f_d=f_h$ it is not possible to predict. The theoretical ability
to  predict, $\theta_c$, depends on how far apart these two
 distributions are in
some sense. We have introduced the {\it parameter} $\theta_e=
1-\theta_c$ that is a measure of this distance and 
represents this predictibility. It is linearly related to the 
 sum of the absolute differences of $f_d$ and $f_h$. 
The {\it training estimate}, the naive
estimate of $\theta_e$, is negatively biased. The 
{\it out of sample estimate}, which is not a true statistic, is positively biased. These estimates can be used to estimate upper and lower bounds on
predictivity. Moreover we have described
these biases and how to modify these estimates to reduce the bias.

One original goal was to explain why the statistic $I$ used in the
Partition Retention method of variable selection is well correlated with
$\theta_e$. We see that $I/n$ is an estimate, with small bias,
 of a parameter $\theta_I$ which is half the sum of the squared differences
of $f_d$ and $f_h$. Consequently $I$ tends to be correlated with predictivity,
and an inequality relating the sum of squares and the sum of absolute 
values yields a simple
estimate of an upper bound on $\theta_e$ from $I$. 
These bounds are useful in telling the analyst how efficient his algorithms
are in squeezing the available information out of the data.

If we examine the chi-square statistic, it converges for large  sample sizes
to a parameter which is a sum  of terms with small probabilities in the
denominator and is not as well related to $\theta_e$ as $\theta_{I0}$.
Moreover, to get reliable results, the sample size has to be sufficiently
large so that there are very few values of $x$ with small frequencies.
The statistic $I$ is not sensative to the effect of small or empty cells.

While theoretical predictivity can not decrease when additional variables are
added to the group to be used for prediction, 
the corresponding decrease in cell frequencies
makes it more difficult to determine an appropriate classification method
and to estimate $\theta_e$.
On the other hand the value of $I$ is degraded by a substantial factor when
a variable with little or no influence is added to the group to be used.Thus
the use of $I$ in Partition Retention tends to find small groups of 
interacting influential variables and to discard noninfluential variables. 

We referred to a previous publication where pattern recognition techniques
were employed to get very good predictions using the variables in several
small interacting groups, each with $\theta_e$ a lttle less than 0.5. 

\section{APPENDIX: INEQUALITY}
It is well known that for real vectors ${\bf x}$,
$\sum x_i^2 \leq (\sum|x_i|)^2$, and we have equality if and only if at 
most one of the components of ${\bf x}$ is not zero. We generalize this
fact with \\ 
Theorem:  If $\sum |x_i|=1$ and $\sum x_i= a$ ,then
$$\sum x_i^2 \leq (1+a^2)/2, $$
and the sum of squares is maximized when $|a|<1$ if and
only if all but two of the components of {\bf x} are 0.

Proof: It is obvious that $|a| \leq 1$. Let $S_1$ be the sum of the
positive values of $x_i$ and $S_2$ the sum of the negative values.  
Let $T_1$ be the sum of the squares of the positive values and $T_2$
the sum of the squares of the negative values. It follows that  
$S_1+S_2=a$ and $S_1-S_2=1$ and thus $S_1=(a+1)/2$ and $S_2=(a-1)/2$. 
Then $T_1\leq S_1^2$ and $T_2\leq S_2^2$. It follows that 
$$ \sum x_i^2=T_1+T_2\leq S_1^2+S_2^2=(1+a^2)/2. $$
Equality is attained if $|a|<1$ when there are at most one positive and one
negative component of ${\bf x}$.


\end{document}